 \documentclass[preprint]{aastex}

\shorttitle{Kepler Stars with Companions}
\shortauthors{Horch et al.}

\begin{document}
\newcommand{\ea}{{\it et al.\ }}


\title{Most Sub-Arcsecond Companions of Kepler Exoplanet Candidate Host 
Stars are Gravitationally Bound}

\author{Elliott P. Horch\altaffilmark{1,5,6,7}, 
Steve B. Howell\altaffilmark{2,6,7}, 
Mark E. Everett\altaffilmark{3,6,7},
and David R. Ciardi\altaffilmark{4,6,7}}

\affil{$^{1}$Department of Physics, 
Southern Connecticut State University,
501 Crescent Street, New Haven, CT 06515}

\affil{$^{2}$NASA Ames Research Center,
Moffett Field, CA 94035}

\affil{$^3$National Optical Astronomy Observatory,
950 N. Cherry Ave, Tucson, AZ 85719}

\affil{$^4$NASA Exoplanet Science Institute, California Institute of Technology,
770 South Wilson Avenue, Mail Code 100-22, Pasadena, CA 91125}

\email{horche2@southernct.edu, 
steve.b.howell@nasa.gov,
everett@noao.edu,
ciardi@ipac.caltech.edu}

\altaffiltext{5}{Adjunct Astronomer, Lowell Observatory.}
\altaffiltext{6}{Visiting Astronomer, Gemini Observatory,
National Optical Astronomy Observatory,
which is operated by the Association of
Universities for Research in Astronomy, Inc., under a cooperative agreement 
with the NSF on behalf of
the Gemini partnership: the National Science Foundation (United States), 
the Science and Technology
Facilities Council (United Kingdom), 
the National Research Council (Canada), CONICYT (Chile), the
Australian Research Council (Australia), Minist\'{e}rio da Ci\'{e}ncia, 
Tecnologia e Inova\~{n}ao (Brazil) and Ministerio
de Ciencia, Tecnolog\'{\i}a e Innovaci\'{o}n Productiva (Argentina).}
\altaffiltext{7}{Visiting Astronomer, Kitt Peak National Observatory,
National Optical Astronomy Observatory,
which is operated by the Association of
Universities for Research in Astronomy, Inc.\ (AURA)
under cooperative agreement with the National Science Foundation. }

\begin{abstract}

Using the known detection limits for high-resolution imaging observations 
and the statistical properties of true binary and line-of-sight companions, 
we estimate the
binary fraction of {\it Kepler} exoplanet host stars. Our
speckle imaging programs at the WIYN 3.5-m and Gemini North 8.1-m telescopes
have observed over 600 {\it Kepler} objects of interest (KOIs) and 
detected 49 stellar companions within $\sim$1 arcsecond.
Assuming binary stars 
follow a log-normal period distribution for an effective temperature range
of 3,000 to 10,000 K, then the  model predicts that the vast
majority of detected sub-arcsecond companions are long period ($P>50$ years),
gravitationally bound companions. In comparing the model 
predictions to the number of real
detections in both observational programs, we conclude that the overall binary 
fraction of host stars is similar to
the 40-50\% rate observed for 
field stars.

\end{abstract}


\section{Introduction}

The {\it Kepler}
 mission has confirmed several hundred exoplanets, and has flagged
thousands of stars as ``Objects of Interest" (KOIs); 
that is, stars exhibiting a 
transit-like event in their light curve. Most of these stars are 
thought to harbor one or more exoplanets, but there will be some 
false positives, caused by either periodic stellar phenomena or the presence of
an unresolved object within the same {\it Kepler} pixel as the object of
interest, such as a background eclipsing binary. 
Determining whether the signals obtained by {\it Kepler}
are caused by
an exoplanet requires a detailed analysis of the light 
curve ({\it e.g.\ }Lissauer \ea 2014 and references therein) 
as well as ground-based follow-up 
observations, including spectroscopy ({\it e.g.\ }Everett \ea 2013)
and high-resolution imaging
({\it e.g.\ }Adams \ea 2012; Howell \ea 2011),
to rule out as much parameter space for false positives as possible.

{ A number of recent papers have discussed the links between occurrence 
and planet properties in relation to single stars and in terms of stellar 
multiplicity. {\it Kepler} data has provided an important impetus
for work in this area. For example, planet occurrence studies looking at the 
general trend of planet radii discovered by Kepler are contained in 
Howard \ea (2012) and Fressin \ea (2013). The latter reference 
summarizes a number of previous and concurrent studies as a function of 
planet radius from Earth-size to giant planets, in particular
concluding that about one in six main sequence stars in the F through
K spectral range has an earth-sized planet with a period of less than 85 days.
Regarding the issue of stellar multiplicity, several 
studies indicate
that the presence of a stellar companion will affect planetary formation
({\it e.g.\ }Xie, Zhou \& Ge 2010; Kraus \ea 2012; 
Parker \& Quanz 2013). 
Wang \ea (2014a, 2014b) 
discuss planet formation and occurrence and the
second reference provides a very good summary of
approximately a dozen previous works on the subject of host star 
multiplicity using a range of observational techniques including 
spectroscopy, adaptive optics,
and lucky imaging.}

The work presented here involves high-resolution imaging using the
technique of speckle imaging. Due to the use of electron-multiplying
CCD cameras in recent years, 
the technique can deliver diffraction-limited images of 
stellar targets over a broad range of stellar brightness 
with relatively high dynamic range. For example, 
stellar companions 
up to 3-4 magnitudes fainter than a target star 
can be seen in the visible range 
within 0.1 arcseconds of the target
star in many cases at the WIYN 3.5-m Telescope\footnotemark{\hspace{1pt}}
for targets as faint as 
13th-14th magnitude (see Horch \ea 2010). 
The upper limit for the separation of components typically observed with
speckle imaging is on the order of 1 arc second; this is usually
set by the field of view of the speckle camera and/or the lack of isoplanicity
at larger separations. For wider separations,
it is possible to search for companions using traditional imaging methods.
Speckle imaging can therefore be useful in learning whether KOIs
have stellar companions 
from the diffraction limit
of a large ground-based telescope (20 to 40 mas) up to $\sim$1 arcsecond.
Given that KOIs lie mainly between $\sim$200 pc
and $\sim$1 kpc of the
Sun, this translates into range of projected separations of a few to 
$\sim$1000 AU. 

\footnotetext{The WIYN Observatory is a joint facility of
the University of Wisconsin-Madison, Indiana
University, Yale University, and the National
Optical Astronomy Observatories.}

The known log-normal period distribution
for solar-type field binaries has its peak at 180 years 
(Duquennoy and Mayor 1991;
Raghavan 2010). In rough terms, this period implies
a semi-major axis on the order of 40 AU, with about two thirds of binaries
having semi-major axes between a few and 200 AU. 
Comparing with the numbers above, this illustrates that speckle imaging of 
stars is therefore an excellent way to learn more about binary 
statistics in this region of the Galaxy in general, 
including the dependence of binary
parameters on spectral type, metallicity, and age. The advent of the 
KOI data sets (Borucki \ea 2011;
Batalha \ea 2013; Burke \ea 2014) represents a group 
of stars where most members 
probably host an exoplanet or system of exoplanets. If
a subsample of KOIs that have confirmed exoplanets and bound stellar
companions can be identified, this would be
 an important tool in understanding
the relationship between binarity and planetary systems.
In this paper, we simulate the observable properties of a sample of KOIs
(in terms of magnitude and separation),
and compare to the speckle imaging data obtained so far
on KOIs. This sets the stage for further work to more rigorously and
systematically identify binaries with exoplanets from the {\it Kepler}
data set.

\section{Observational Sample}

Since 2008, our group has been taking speckle observations of 
KOIs with the Differential Speckle Survey Instrument (Horch \ea 2009). 
The camera records speckle images in
two filters simultaneously, so that each observation results in two 
diffraction-limited image reconstructions of the target. In the first
channel of the instrument, we have always used a 692 nm filter with 
width of 40 nm. However, in the second channel, 
we have used both an 880 nm with filter width of 50 nm 
and a 562-nm filter of width 40 nm. While the 562-nm filter gives superior
resolution images (owing to its shorter wavelength), 
we find that the 880-nm filter often gives images
with a larger dynamic range, and is therefore sensitive to the detection
of fainter companions. In a small number of cases, high-quality data only
exist at 692 nm
for the target because the data in the other channel were degraded by 
a scattered light problem in the instrument that has since been resolved.
As such, the only filter where all targets observed have data is 692 nm,
and this represents our most comprehensive data set.

The basic observing strategy has been to obtain data on as many KOI targets 
as possible on each observing run, although we have tended to observe
targets whose planetary candidate (or at least one of the planetary candidates)
is ``Earth{-sized},'' {\it i.e.\ }having a derived radius less than about 3
Earth radii. We have rarely repeated observations on
these stars, even those with discovered companions, up to this point.
For the vast majority of cases, this means we have only one observation
in 692 nm and one other observation in one of the other two filters 
mentioned above.
We have found sub-arcsecond companions to 49 {\it Kepler} stars
out of a grand total of over 600 observed at present, 
combining results from 
both the WIYN 3.5-m Telescope 
at Kitt Peak and
the Gemini North 8.1-m Telescope on Mauna Kea. 
{ As we will discuss further in Sections 4 and 5, 
a number of these are expected 
to be false positives; current information on the CFOP 
website\footnotemark{\hspace{1pt}}
 indicates
that of these stars, 74 have been judged to be false positives (12\%), 
eleven of
which are found to be binary in our speckle observations. This latter number
represents 22\% of those stars detected as binary so far.}

\footnotetext{See {\tt https://cfop.ipac.caltech.edu/home/}.}

Since the speckle data that we have is in two filters, it would in principle
be possible to put the components of any binary or multiple stars detected
onto the H-R diagram to test whether a common isochrone is consistent
with the positions of the stars. However, 
the speckle analysis of WIYN data has been shown to give magnitude 
differences between
components with uncertainties in the 0.1 to 0.2 magnitude range for a wide 
range of component brightnesses (see {\it e.g.\ }Figure 7 in Horch \ea 2010). 
Therefore, the color information that we have at present is uncertain
to 0.14 to 0.28 magnitudes, depending on signal-to-noise ratio and other 
factors. As such, in the vast majority of cases, we do not have sufficient 
leverage on the component colors to attempt an analysis of this type at present.
Likewise, with a sufficient number of observations over a period of years,
it would be possible in principle to detect orbital motion, or common proper
motion, based on data obtained from the speckle camera. However, we do
not yet have the data necessary for this analysis because of the observing
strategy mentioned above; only a handful of stars have multiple observations.
We show in Table 1 details of the observations obtained at WIYN and Gemini 
North. 
More information on the speckle observing
process can be found in Howell \ea 2011 and Horch \ea 2011 (WIYN) and 
Horch \ea 2012 (Gemini). A more detailed analysis of these systems including
final relative astrometry and photometry, as well as placement of
components on the H-R diagram, will be forthcoming when sufficient follow-up
data exist.

We have excluded Gemini observations of three KOIs in this study, namely
those of KOI 98, KOI 284, and KOI 2626. 
The first two of these objects were known
to be double from WIYN observations, and the Gemini observation was made to
confirm the earlier result and compare data quality directly with 
the WIYN observations.
These objects only appear in the
figures presented here of WIYN data and in the WIYN statistics displayed 
in Table 1.
We have observed the third object, KOI 2626, on three occasions at Gemini, but
this object is removed from consideration here as those
observations were to confirm Adaptive Optics results that had already
been obtained at the Keck Observatory. This object was never observed at WIYN,
and does not appear anywhere in the data presented here.

In Figures 1 and 2, we show further properties of the observed sample for
both telescopes. In Figure 1, we show the placement of objects in the 
{\it Kepler} field; this illustrates that the samples are consistent with a 
random distribution in terms of sky position. In Figure 2, we show 
the {\it Kepler} magnitude of the stars as a function of estimated distance;
drawing upon Huber \ea 2014, a distance modulus is obtained from the 
spectral type implied from the
effective temperature and known surface gravity of the star
and inferring an absolute magnitude from that. { This does not account for
the binarity or multiplicity of some of the stars in the sample, 
which would affect the 
estimated distance and stellar properties to some degree depending
on the magnitude difference of the components; the distances obtained
are intended only to show that, in rough terms, the sample of observed stars
is similar to the simulation results.}
Finally, in Figure 3, we show the surface gravity as a function of
effective temperature; this shows that while the sample spans a range
in effective temperature from about 3,000 to 10,000 K, it is dominated by
dwarfs that have near-solar values in both quantities.

\section{Method}

We wish to study the number of bound versus the line-of-sight
companions that will be detected by the DSSI instrument when looking
at stars randomly selected in the {\it Kepler} field.
The vast majority of {\it Kepler} stars are in a distance range of 
roughly 200 to 1000 pc
relative to the Solar system (corresponding to a diffraction-limited
separation of 10 to 50 AU at WIYN, and 4 to 20 AU at Gemini North).
The range of Galactic latitude
and longitude appropriate for the {\it Kepler} field is
$5.5 \le b \le 21.48^{\circ}$ and
$68.1 \le l \le 84.5^{\circ}$, respectively.

We have used the TRILEGAL galaxy model (Girardi \ea 2005) 
to construct simulations
of star counts in the {\it Kepler} 
field of view. Ten randomly selected pointings
within the {\it Kepler} field were used; these are shown in Figure 4.
This produced 10 lists of stars, which
were then combined to give better statistical results of the properties 
of the stars in the entire field. Each of the ten simulations had 
a field of view
of one square degree, but the simulations were run with the binary parameters
turned off. Since for this study we required detailed information regarding
the companion stars and their orbital properties, these were added after 
the fact as follows.
From the TRILEGAL output, we constructed a distance-limited sample with
maximum distance from the solar system of 1300 pc. In order to study 
only stars like those observed, we excluded stars with effective temperatures
less than 3,000 K and greater than 10,000 K, and also required that 
log($g$) was between 3.3 and 4.7, although, since the observed sample had 
only a small percentage of higher-temperature  stars
(as can be seen from Figure 3),
we removed, at random, 50 percent of those stars with effective temperature
greater than
7,000 K. 

As this was a distance-limited
sample overwhelmingly dominated by solar-type stars, it is reasonable to
add companions according to the known statistics of 
the field population of binaries (Duquennoy and Mayor 1991; 
Raghavan et al 2010). Specifically, we populate the stars in the
sample with companions at the rate of { 46}\%. We find a mass ratio for
each system by utilizing the mass-ratio distribution found in Raghavan 
\ea 2010 (specifically, Figure 16 in that work).
The mass of the primary is known from the TRILEGAL output, so the secondary 
mass is then calculated.
From the 
mass-luminosity relation of Henry \& McCarthy (1993), 
these masses can be converted
into absolute $V$ magnitudes. 
Using the distance, an apparent magnitude
can be calculated as well as a magnitude difference for the binary components.
Finally, we convert this magnitude difference at $V$ to the speckle
692-nm filter by estimating the spectral type of the primary and 
secondary from the mass values and using the known filter transmission curves.

In the case of binary stars, we select a period according to the 
Duquennoy \& Mayor (1991) log-normal period distribution, and an eccentricity
for the orbit using the information in the same paper. We select random
values for the cosine of the inclination ($\cos i$), 
ascending node ($\Omega$), and 
the angle in the true orbit between the line of nodes and the semi-major 
axis ($\omega$), and time of periastron passage ($T$). Finally, 
we determine the semi-major axis in AU from the masses and the period and 
convert this to arcseconds using the distance. With
the seven orbital parameters in hand, we can then compute the ephemeris 
position angle and separation for a randomly chosen epoch of observation. 

We then test whether a companion would be detected for each star in the
sample 
using the camera and telescope combinations from our work (whether
single or double). 
In order to make this 
determination, we first select stars that have an apparent magnitude 
brighter than the detection limit for the telescope in question
(14.5 at WIYN, 16.5 at Gemini North). We assume that single stars
would be seen as single by DSSI, but for binaries, we next
apply an average contrast limit curve for WIYN and
Gemini,
that is, a curve of the maximum observable magnitude difference
as a function of separation from the central star. 
The process for making detection limit curves for 
{\it Kepler} has been described in {\it e.g.\ }Howell \ea (2011) and 
Horch \ea (2011),
but briefly, we use the reconstructed 
({\it i.e.\ }diffraction-limited)
images from the speckle data in order to estimate
such curves for all stars observed. 
We determine the values of all the local maximum ``sky'' pixels
     within a set of concentric annuli centered on the target star. 
     Detection limits for a point source at a given radius are
     calculated using the appropriate annulus and are set to the mean of
     the maxima plus 5 times the standard deviation of the maxima.
     If its magnitude difference is less than the value of this curve
     for the separation of the system, then a companion is considered
     to be detectable. 

{ DSSI is essentially a magnitude-limited instrument
at each telescope, sensitive to targets brighter than $V \sim 14.5$ 
at WIYN and 16.5 at Gemini, although these boundaries are influenced 
somewhat by observing conditions. Because of this fact, we anticipate that
some binaries where both stars' magnitudes lie below the detection threshold
will be nonetheless detectable due to the combined light, thereby
creating a potential bias due to faint, primarily small--magnitude-difference
pairs. However, because the simulations are first distance-limited,
and then binaries are added to this entire sample before imposing the
detection limit of the camera, the simulation results also reflect this
bias. For example, in the WIYN simulations, about 15\% of binaries
detected had both primary and secondary magnitudes below the detection
threshold, and for Gemini, the result was about 6\%. (It is lower in
this case due to the fact that the sample is dominated by G dwarfs within
$\sim$1000 pc, and generally 
have apparent magnitudes well above the Gemini detection
limit for the relevant range of distances.) 
Therefore, we have accounted for the observational bias
by in effect building the same bias into the simulations.}

The result of this simulation scheme is shown in Figures 5 and 6 for 
both WIYN and Gemini North. Figure 5 shows the {\it Kepler} magnitude of the
sample as a function of distance and may be directly compared with Figure 2,
while Figure 6 shows a plot of log($g$) versus effective temperature, and
may be compared with Figure 3. Only one of the 10 pointings in the {\it Kepler}
field was used to make Figures 5 and 6 in order to keep the figures clear;
In Figure 7, we show histograms of effective temperature and $\log(g)$
for the observed sample of stars at WIYN
along with histograms
of the complete ``observable'' sample from the WIYN simulation, including
all 10 pointings. (Plots for Gemini appear very similar, but with many
fewer observations and much lower statistical significance.)
On this basis, we judge these simulated samples to be
sufficiently close to the actual observed samples in magnitude, distance,
and stellar make-up to be useful in predicting the percentage of detected
companions at each telescope with the speckle instrument.
Given this approach,
the periods for the detected binaries in the Gemini simulation ranged 
from 17 to 376,000 years, with a median value of 970 years. For the 
WIYN simulation, the minimum period was 29 years, 
the maximum was  49,400 years,
and the median value was 1600 years.

To determine the frequency of optical doubles ({\it i.e.\ }line-of-sight
components), we assigned random positions within the field
to the stars in the output
file of each TRILEGAL simulation, and then computed the distance on the sky
between these stars and each of the previously identified observable stars.
For stars that had a separation of less than 1.2 arc seconds, we computed
a magnitude difference, and use the contrast limit curves to determine if 
the object would be seen as double when observed with the speckle camera.

\section{Results}

Using the simulations described above, we can now compare to the observed
data in terms of the properties of detected components.
In Figures 8 and 9, we show the magnitude difference of all detected 
and simulated
binaries as a function of separation. We also plot the average
detection limits at 692 nm as a function 
of separation used for this study for WIYN and Gemini North respectively.
For the observed data,
the percentage of KOIs where companions were discovered
at WIYN is $7.0 \pm 1.1$\% (41 of 588 targets observed), whereas for 
Gemini it is $22.8 \pm 8.1$\% (8 of 35 targets observed).

Note that in Figure 8 there are seven systems that are above the contrast
limit curve. The curve we have selected for the analysis here is an
average of several obtained from unresolved objects 
that have apparent magnitudes comparable to the normal range of 
{\it Kepler}
objects observed at WIYN, between 11th and 14th magnitude. However, some 
{\it Kepler} stars are significantly 
brighter than this, and so would have much higher 
signal-to-noise than the typical {\it Kepler} observation.  In addition, the 
contrast limit curve will be higher for objects taken in better 
seeing conditions. It is not uncommon 
to detect companions at magnitude differences of 5
at WIYN for brighter targets
observed in good seeing 
(see {\it e.g.\ }Horch \ea 2011). So, while there is some variation in the 
detection limit curves for individual observations, the curve shown is 
a reasonable average for {\it Kepler} observations. Without the seven 
detections shown above the detection limit curve that is drawn (and all
seven represent sources brighter than magnitude 12.2), then
the observed rate of companion detection at WIYN would be $5.8 \pm 1.0$\%.
However, this number is likely to be an underestimate of the true 
WIYN detection rate for the detection limit curve shown, 
as some stars would have been
observed in poor conditions where fainter companions below the
detection limit curve would still be missed.
In contrast, the same situation does not exist at Gemini, since 
the larger telescope 
aperture puts nearly all {\it Kepler} stars observed to date 
into the high
signal-to-noise regime.

Also in Figures 8 and 9, we show the simulation results obtained as described 
in Section 2, where the detection limit curves shown in 
Figures 8 and 9 are assumed. We find that, in the case of WIYN simulations, the 
percentage of detected companions predicted is 
$7.8 \pm 0.4$\% (451 of 5,745 trials). Of these 
companions, 96\% are predicted to be bound companions (with 
the remaining 4\%
being optical doubles). In the case of Gemini data, the rate of
companion detection is predicted to be 
$19.7 \pm 0.4$\% (2,148 of 10,879 trials), 84\% of which are
predicted to be gravitationally bound systems, 94\% of systems
with $\Delta m < 5$ are gravitationally bound.

In both cases, we find reasonably good agreement with the 
observed rate of
detections when assuming the 46\% number for stars with companions from
Raghavan \ea 2010. (That is, we are using 
$7.1 \pm 1.1$\% observed for WIYN versus
$7.8 \pm 0.4$\% predicted and $22.8 \pm 8.1$\% observed for Gemini versus
$19.7 \pm 0.4$\% predicted.) { These numbers give confidence that
our detection limits are well-understood.
In} Figures 8 and 9, it is interesting to note the segregation 
of the two types of companions
particularly in the Gemini simulation, with most line-of-sight companions
being at larger separations and higher magnitude differences. In 
contrast, the 
bound stellar companions cluster toward smaller separations, with typical
semi-major axes of $\lesssim$90 AU (periods less than $\sim$700 years).

{ To make a preliminary statement regarding the binarity of exoplanet
host stars, we first remove from the above statistics those 
objects judged to be false positives as of the present. This gives us
the cleanest possible sample of exoplanet candidate host stars with which 
to work.
In this case, the WIYN detection rate is reduced to $6.2 \pm 1.1$\% (32
of 518 stars), slightly below
that of the simulations, while the Gemini detection rate remains 
fairly constant and consistent with the simulations, at $20.0 \pm 8.2$\%
(6 of 30 stars).
We have redone the simulations, changing the input rate of companions
to see what effect that would have on the final prediction for 
companion detection.
From this, 
we can estimate that
the companion star fraction of this ``clean" sample at WIYN 
is $37 \pm 7$\% 
at present. A similar study of Gemini data resulted in 
an estimate of the companion star fraction of $47 \pm 19$\%. These 
numbers bracket the 40-50\% range believed to be the
case for field stars.}

\section{Discussion}

\subsection{General Comments}

The Duquennoy \& Mayor log-normal period distribution for binary stars 
has its peak
at a period of 180 years for G type stars; there is less information
in the literature 
about other spectral types. Nonetheless, if the distribution is 
similar for main sequence spectral types A through M, 
the data indicate that, even at 
Gemini with its much more sensitive detection limits, most 
observed close companions
will be gravitationally bound. 

It is estimated that, in general, over 90\% of KOI stars do indeed harbor 
transiting exoplanets, {\it i.e.\ }the false 
positive rate is thought to be under 10\% (Morton \& Johnson 2011; 
Fressin \ea 2013; Santerne \ea 2013), though it is 
generally believed that the presence of a companion increases the 
false positive probability due to the possibility of the companion 
being a background eclipsing binary star. Likewise, the identification 
of multiple planet candidates and/or the orbital period of a planet candidate 
also can decrease the false positive rate (Lissauer \ea 2014). 
For truly bound companions
detected here
it is not clear that the false positive rate should increase, 
since an eclipsing binary bound to the 
KOI star would contribute more light in general than a background system and
would therefore be more likely to have deeper transit-like events in the 
{\it Kepler} data stream. It would be more 
easily recognized as a false positive.
Combining this line of thinking 
with the results obtained here, 
it implies that the sample of detections shown in Figures 8 and 9 is
mainly comprised of binary systems where one of the stars harbors an
exoplanet. Since the binary statistics derived here come from 
those of the known
field population, it 
would appear that, for the full range of separations and periods 
to which we are sensitive,
the binary fraction of stars that have exoplanets is overall
roughly consistent with 
that of the
field population.
Otherwise our observed detection
rates would not match those of the simulations. 

\subsection{Further Vetting}

Since their identification as KOIs, 11 of our 49 discoveries have been
judged to be false positives (see Batalha \ea 2013 and Burke \ea 2014), 
while 12 other targets have been identified
as multi-planet candidate systems (see {\it e.g.\ }Lissauer \ea 2014),
and 6 have been validated as exoplanet hosts, two with multi-planet
systems (see {\it e.g.\ }Marcy \ea 2014 and 
Rowe \ea 2014). 
While we do not yet have sufficient 
speckle data in hand to study the placement of the components of these
systems further on the H-R diagram,
we can at least note the positions of these objects 
in a plot like Figure 8 or 9 and investigate the implications of this
considering where line-of-sight and bound companions are expected to
dominate the sample. This is shown in Figure 10.
We see that the two false positives from the Gemini list
and five from the WIYN list do have positions that put them in the region
populated by light-of-sight companions, whereas 0 Gemini and 4 
WIYN false positives are in a region dominated by bound companions.
In contrast, eight out of twelve of the multi-planet candidate systems 
occur in the
region of separation less than 0.6 arc seconds and magnitude 
differences less than 2.5.
Only one of the validated systems lies in a region where more line-of-sight
companions are expected.
The sample size is still very low, so firm conclusions cannot be made;
however, at this stage there is no obvious inconsistency, as 
{\it e.g.\ }one would expect false positives to be generated 
by background stars in 
many cases.

If the known false positives are discounted, then we are left with a sample
of 38 KOIs with companions detected with speckle imaging, six of which have
already been validated as hosting exoplanets. While it is true
that it is still possible for some of these systems to be judged to be false
positives in the future, the false positive rate is not likely to
be worse than
the standard 10\% number discussed above, which would imply
that roughly 35 or more of the sample are gravitationally bound binaries that
host exoplanets.

\subsection{Planet Radius}

The speckle observations and companion modeling indicate that any star
detected within $\sim$1 arcsecond is almost certainly 
a bound companion star.  If a
target star is blended with another star (bound or line-of-sight), the
true planet radius is larger than the derived planet radius because the
observed transit depth is diluted by the companion star.   In general,
we do not know around which star the planet is orbiting.
If the primary star hosts the planet, then for an equal brightness
binary (assuming no color difference), 
the planet size is underestimated by no more than a factor of
$\sqrt{2}$ and as the binary ratio (and hence, relative brightness) increases,
the observed planet radius asymptotically approaches the true planet
radius. This is shown in the blue line of Figure 11.

If we assume that the companion star is indeed bound as indicated by the
simulations above, then the secondary star must be smaller than the
primary star ({\it i.e.\ }a lower luminosity indicates a smaller star). If the
planet orbits the secondary star, the planet can be significantly larger 
than anticipated because the secondary star is heavily diluted by the
primary star {\it and} the secondary star is smaller than the primary star
({\it i.e.\ }the stellar radius assumed if the planet orbited the primary
star).  As an example, in the red line in Figure 11, we have calculated the
change in the derived planet radius for a G0V primary star with a bound
companion of some stellar type, but it is the companion that hosts the
planet.  Correcting for the transit dilution and the smaller stellar
radius of the secondary star, the true planet radius can be 1.4-8.0
times larger than the planet radius derived from the blended photometry
and assuming the planet orbits the primary star: the fainter the
companion, the larger the planet actually is.  There is, of course, a
limit such that an observed transit depth cannot be mimicked by a
star if the star is too faint and/or the transit depth is too deep.

We have taken the known effective temperatures, magnitude differences,
and planet radii (as appearing on the CFOP website) 
of the 49 double stars
in our sample and estimated spectral types for the secondary star
using information in Schmidt-Kaler (1982). Then, using curves like Figure 11,
we have estimated the factor by which the true planet radius is larger 
than that derived from the transit data { if the planet were to orbit 
the secondary
star}. The result is that 28 of 41
WIYN discoveries (68\%) have planet radii that remain well below the threshold
of a late M-dwarf even if orbiting the secondary, 
and all 8 Gemini discoveries (100\%) remain
below the stellar threshold in radius. { In the WIYN sample, 9 of the
41 stars are now judged to be false positives; five of these are in
the group of 13 stars that do not remain below the stellar radius threshold
in this exercise, a significant overlap.}
We conclude that it is not possible
to explain the majority of transits by suggesting that the secondary is
an eclipsing binary star.

\subsection{Possible Suppression of Small-Separation Stellar Components}

The study of Wang \ea (2014a) concluded that there is a 
suppression of stellar companions inside 20 AU for exoplanet host 
stars.
In the range from 20 to 85 AU, the data were less clear-cut,
and above 85 AU, the binary fraction appeared to be consistent with 
that of the field population. 
The work here is mainly sensitive to separations above 20 AU since the vast
majority of objects detected in the simulations have semi-major
axes greater than this value (only 
32 stars of 1796 detections in the Gemini simulation 
and 4 of 538 in the WIYN
simulation have a semi-major axis less than 20 AU.) However, 35.2\% of
detected binaries in the Gemini simulation and 16.4\% of those detected
in the WIYN simulation have semi-major axes in the
range of 20 to 85 AU. If there were a reduction of {\it e.g.\ }50\% 
in stellar companions in this range,
then the overall predicted Gemini detection rate would be reduced 
only to 16.8\%
while the overall WIYN rate would drop to 7.6\%. 

Both of these values are still
consistent with the observed values at present, but
we note that objects with semi-major axes less than 85 AU predominantly
occur at observed separations less than 0.2 arc seconds (74 of 184 objects 
for the WIYN simulation and 657 of 1139 objects in the Gemini simulation
have both observed separation less than 0.2 and semi-major axis less than
85 AU). Thus, if there were a suppression of binaries with $a < 85$ AU,
this could be seen in a relative lack of detected components at separations
less than 0.2 arc seconds. For WIYN, the simulation results predict that
if no such suppression exists, then 41\% of detections should be made
at separations less than 0.2 arc seconds. 
{ Using our ``clean'' sample from the Section 5.3, out of 32 binaries at
WIYN,}
we would therefore expect ${ 13} \pm 4$ detected below 0.2 arcseconds,
whereas for our sample 7 have been detected. For Gemini, the simulation
indicates that 53\% of companions detected should have separations below 
0.2 arcseconds, or in a sample of { 6}, about ${ 3} \pm 2$, 
whereas only one object so far
as been detected in this category. The sample sizes preclude definitive
statements at this point, but clearly with continued observations of the
KOI targets, there is an opportunity to investigate this important range of
semi-major axes.

{ A second paper by Wang \ea (2014b) 
indicates a potentially much broader but weaker 
suppression of stellar companions for exoplanet host stars,
out to separations as large as 1500 AU. These results are based on 
a combination of adaptive optics and radial velocity observations. Our
results include this larger range of distances, but neither the Wang
et al result nor our work have small enough uncertainties at this stage
to be definitive. While Wang \ea break their results
down by physical separation and we do not, our results
for the value of the companion star fraction
quoted in Section 4 for the ``clean'' exoplanet
host star sample 
are easily less than 2-$\sigma$ 
from their number at 100 AU,
and our number is
consistent with the Wang
\ea result at 1000 AU.
In addition, the possible lack of small separation components in our case
is broadly in line with their findings, though it is difficult to 
quantify at this stage.
We conclude} that
high-resolution imaging techniques
{ (speckle imaging, lucky imaging, 
and adaptive optics)} represent an extremely { important} way to detect
small separation companions and assess the true binary fraction of 
exoplanet host stars. { Of these techniques, speckle imaging at 
Gemini offers the highest spatial resolution (20 mas), and will overlap
to the greatest extent with radial velocity studies.}
Finding a companion to a KOI star 
typically corresponds to the identification of a 
binary host star with one or more exoplanets. 
Our continued observations of {\it Kepler}
and other exoplanet host stars will provide high-precision magnitudes and 
colors of stellar components, eventually allowing us to use isochrone fitting
to place the stars on the H-R diagram, 
as discussed in Davidson \ea (2009).
Such analyses can yield mass information of the components, 
and if one component is evolved, to estimate the age of such systems.

\section{Conclusions}

We have simulated the detection process for speckle imaging 
of {\it Kepler} Objects of Interest 
and compared the rate of companions predicted
with that found so far in real observations. 
We find that the real rates of companion
detection are, within uncertainties, the same as the simulations for 
two different speckle observing situations, the WIYN telescope and 
Gemini North.
The simulations incorporate the TRILEGAL galaxy model to generate lists
of stars and their properties in the {\it Kepler} field. After 
a distance-limited subsample of these objects is constructed, 
the known statistics concerning binarity among stars near the Sun 
is added.
The simulations predict that the very large majority of sub-arcsecond
companions will be physically bound to the {\it Kepler} star.
This result suggests that, over the separation range to which we are 
sensitive, exoplanet host stars have a binary
fraction consistent with that of field stars. Our speckle imaging program has
identified a sample of 
candidate binary-star exoplanet systems in which only a modest number of 
false positives are likely to exist.

\acknowledgments
We thank the {\it Kepler} 
Project Office located at the NASA Ames Research Center
for providing partial financial support for the upgraded DSSI instrument.
It is also a pleasure to thank Steve Hardash, Andy Adamson, Inger Jorgensen,
and the entire summit crew for their assistance at Gemini,
as well as Charles Corson and the team of observing assistants
at WIYN for all of their help during our runs over the last few years.
We also thank the anonymous referee for her/his comments that 
have helped to significantly improve the paper.
This work was funded by the {\it Kepler} Project Office.



\begin{deluxetable}{lcc}
\tabletypesize{\scriptsize}
\tablewidth{0pt}
\tablenum{1}
\tablecaption{Basic Properties of the Observed Data Set}
\tablehead{
\colhead{Parameter} & 
\colhead{WIYN} & 
\colhead{Gemini} \\

}
\startdata

Total Observations & 682 & 42 \\
Stars Observed & 588 & 35 \\
Average {\it Kepler} Magnitude of Sample & 12.85 & 13.04 \\
Companions Detected & 41 &  8 \\
Stars with Multiple Observations & 60 & 2 \\
Stars with 880-nm Filter Data & 453 & 35 \\
Stars with 562-nm Filter Data & 135 & 0 \\

\enddata

\end{deluxetable}

\clearpage

\begin{figure}[tb]
\plottwo{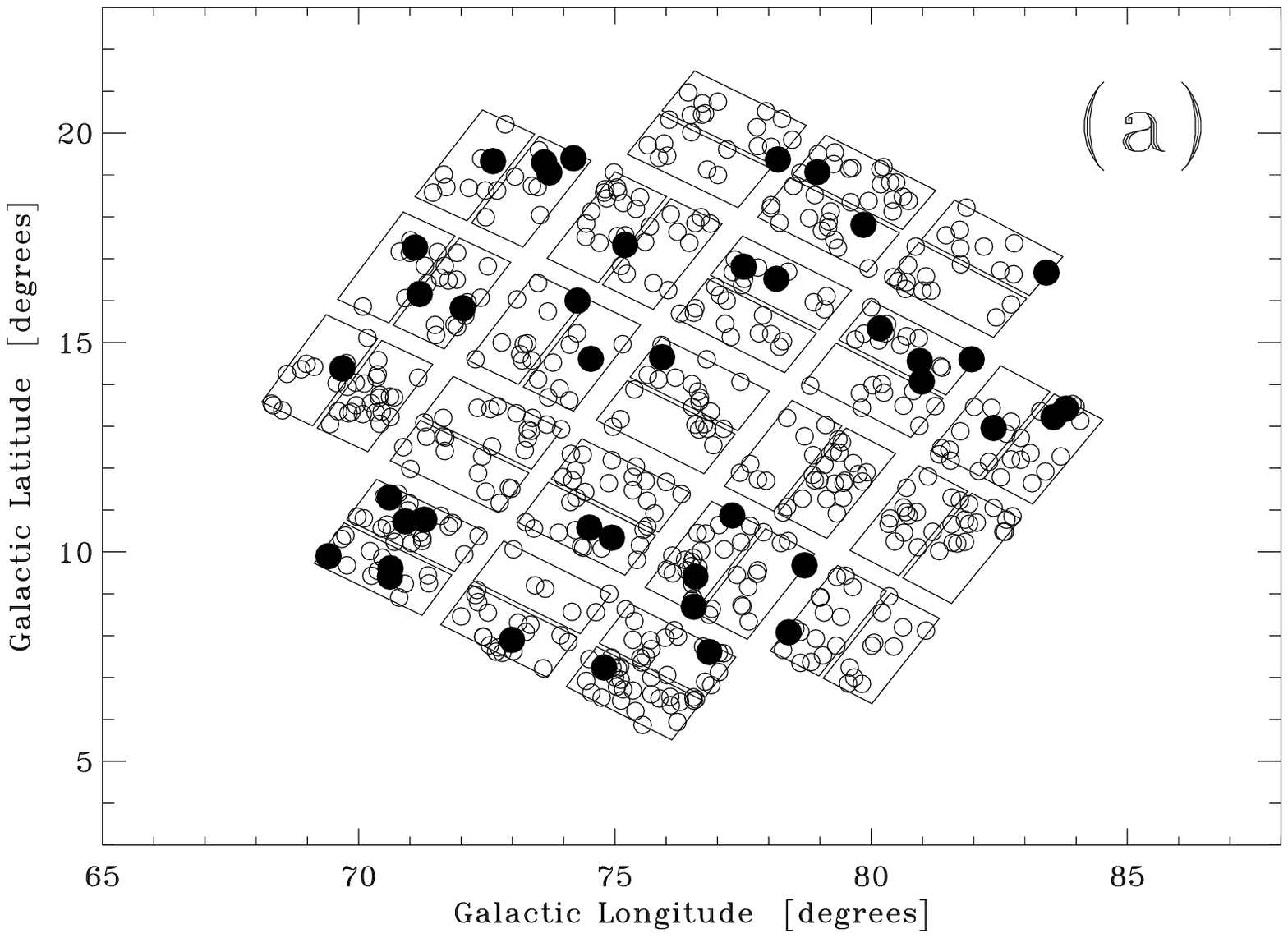}{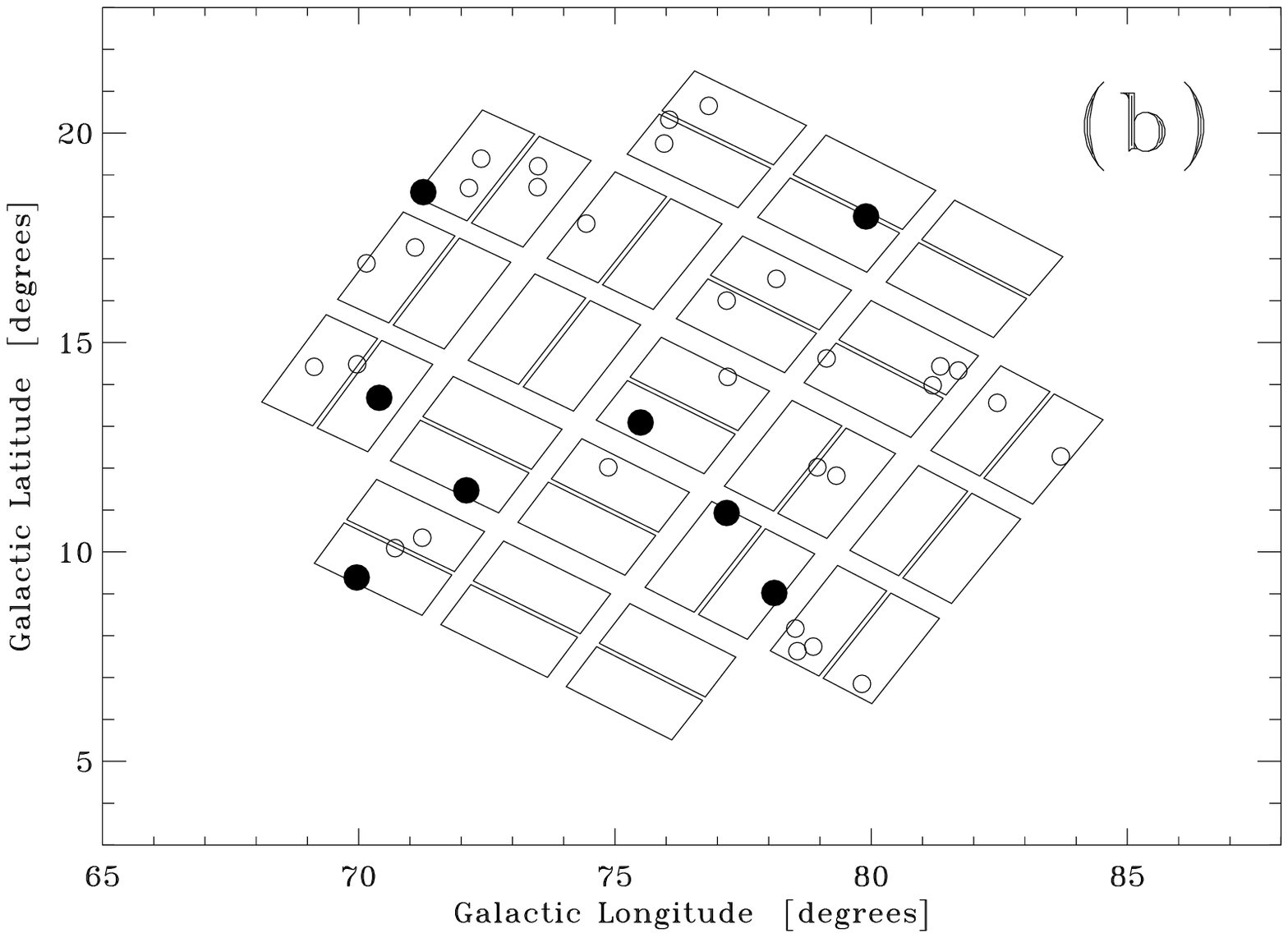}
\caption{
Location in galactic coordinates for the stars observed at (a) WIYN, and
(b) Gemini. Open circles represent stars not found to have companions 
from the speckle observations, and filled circles represent stars where 
a companion has been detected. 
The outline of the {\it Kepler} CCDs is shown by rectangular shapes drawn
with solid lines.
}
\end{figure}

\begin{figure}[tb]
\plottwo{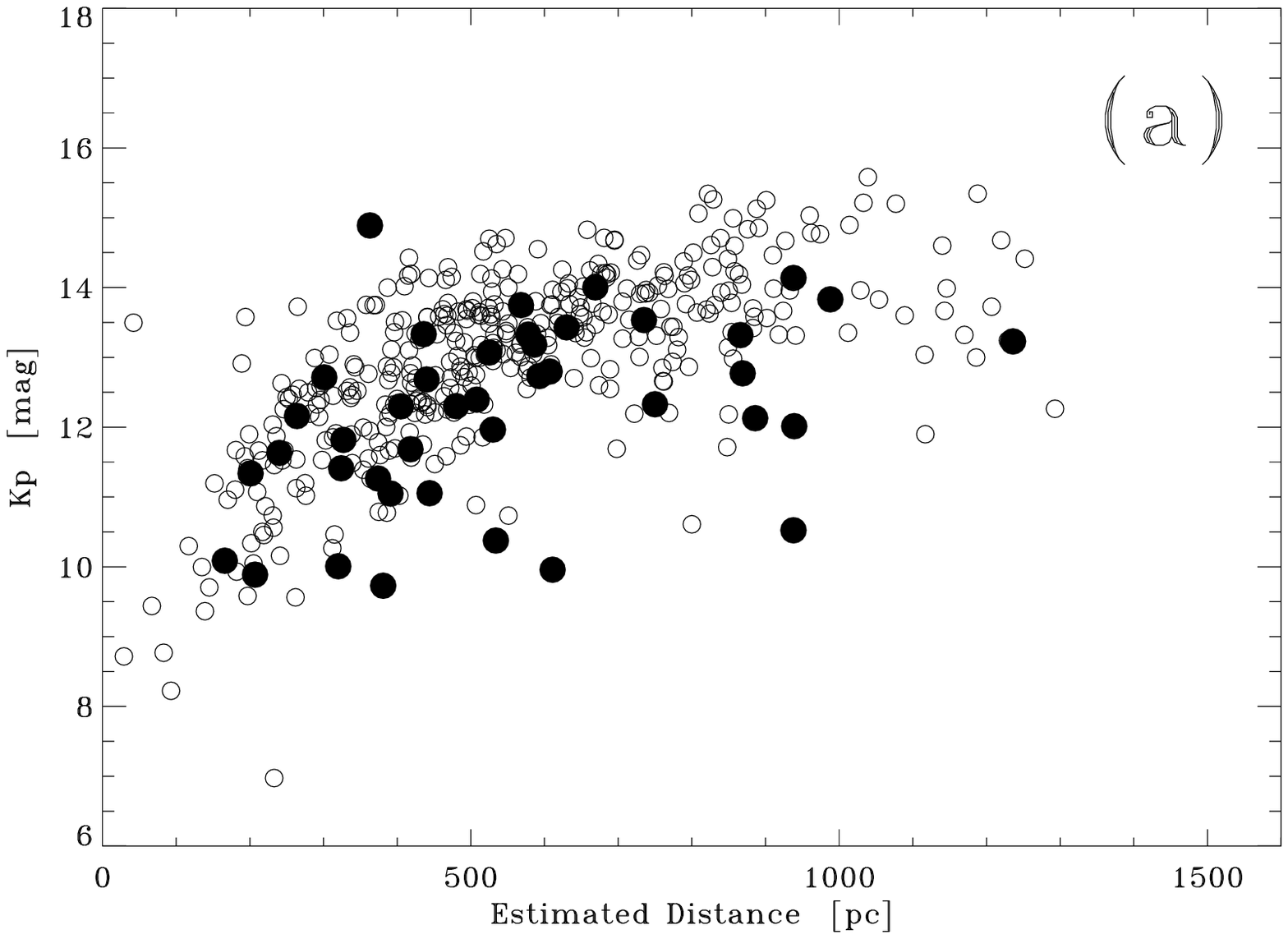}{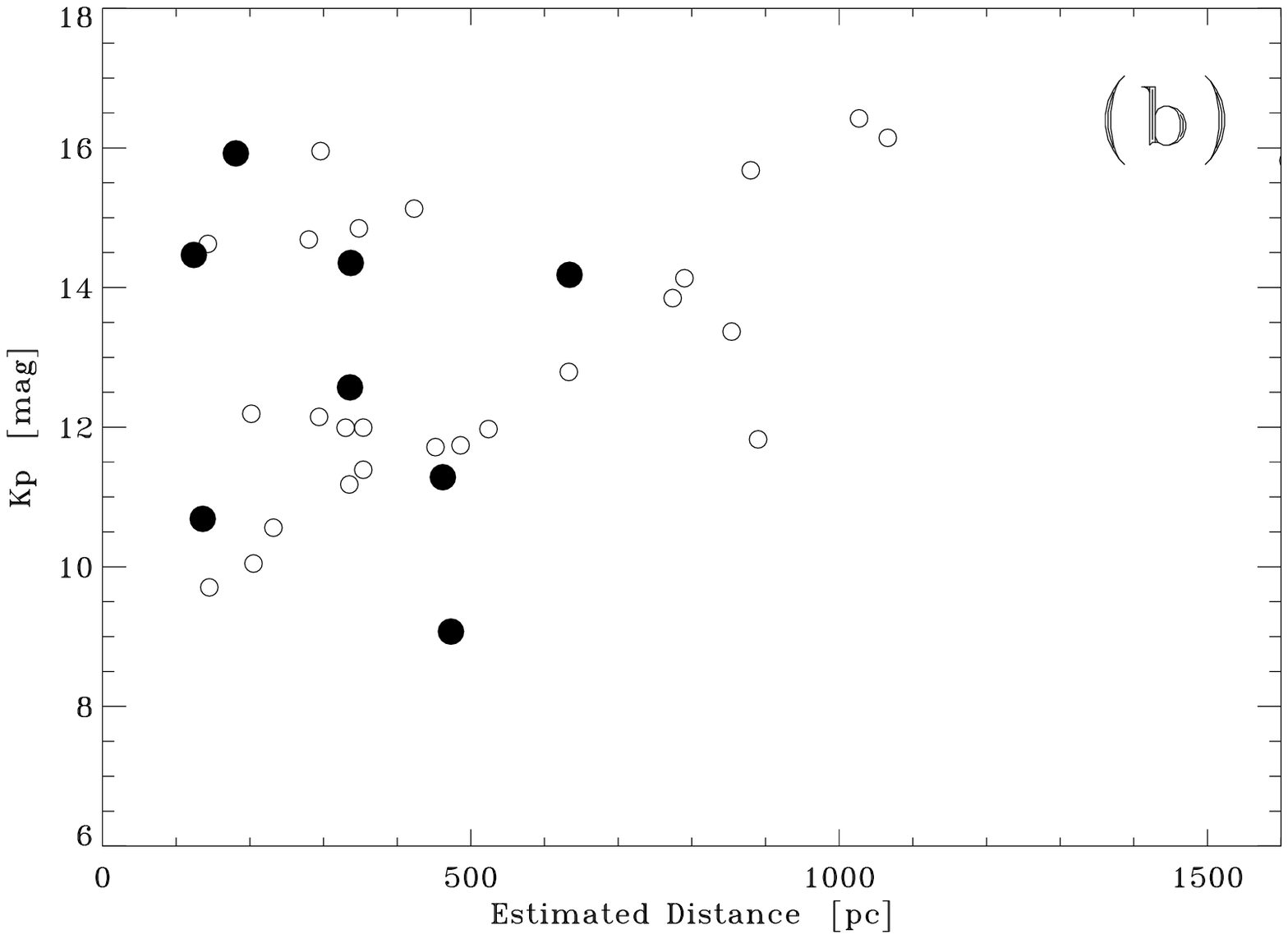}
\caption{
{\it Kepler} magnitude as a function of estimated distance for the sample of
stars observed at (a) WIYN and (b) Gemini. Open circles represent stars
with no detection of a companion, and filled circles represent stars with 
detected companions.
}
\end{figure}

\begin{figure}[tb]
\plottwo{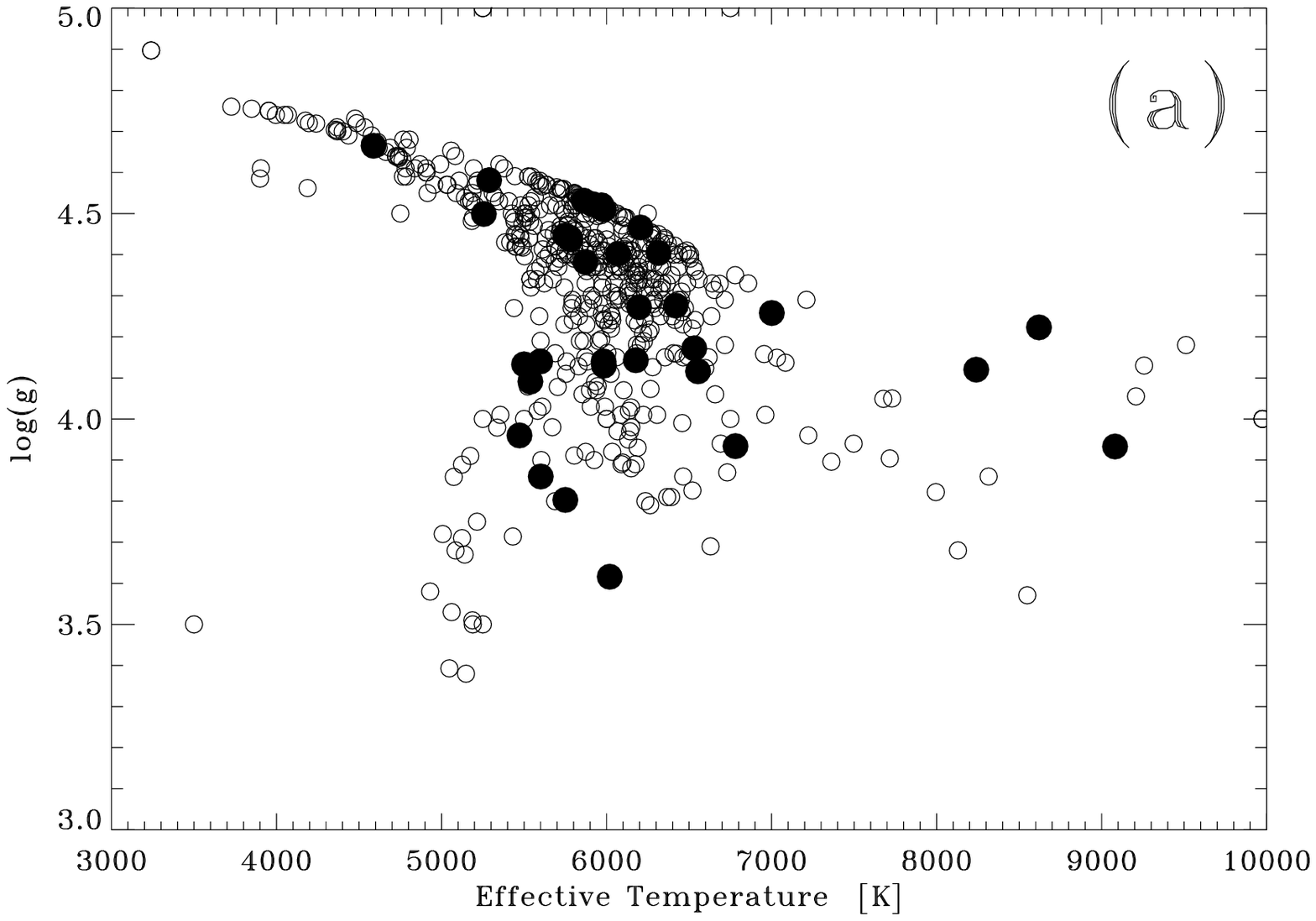}{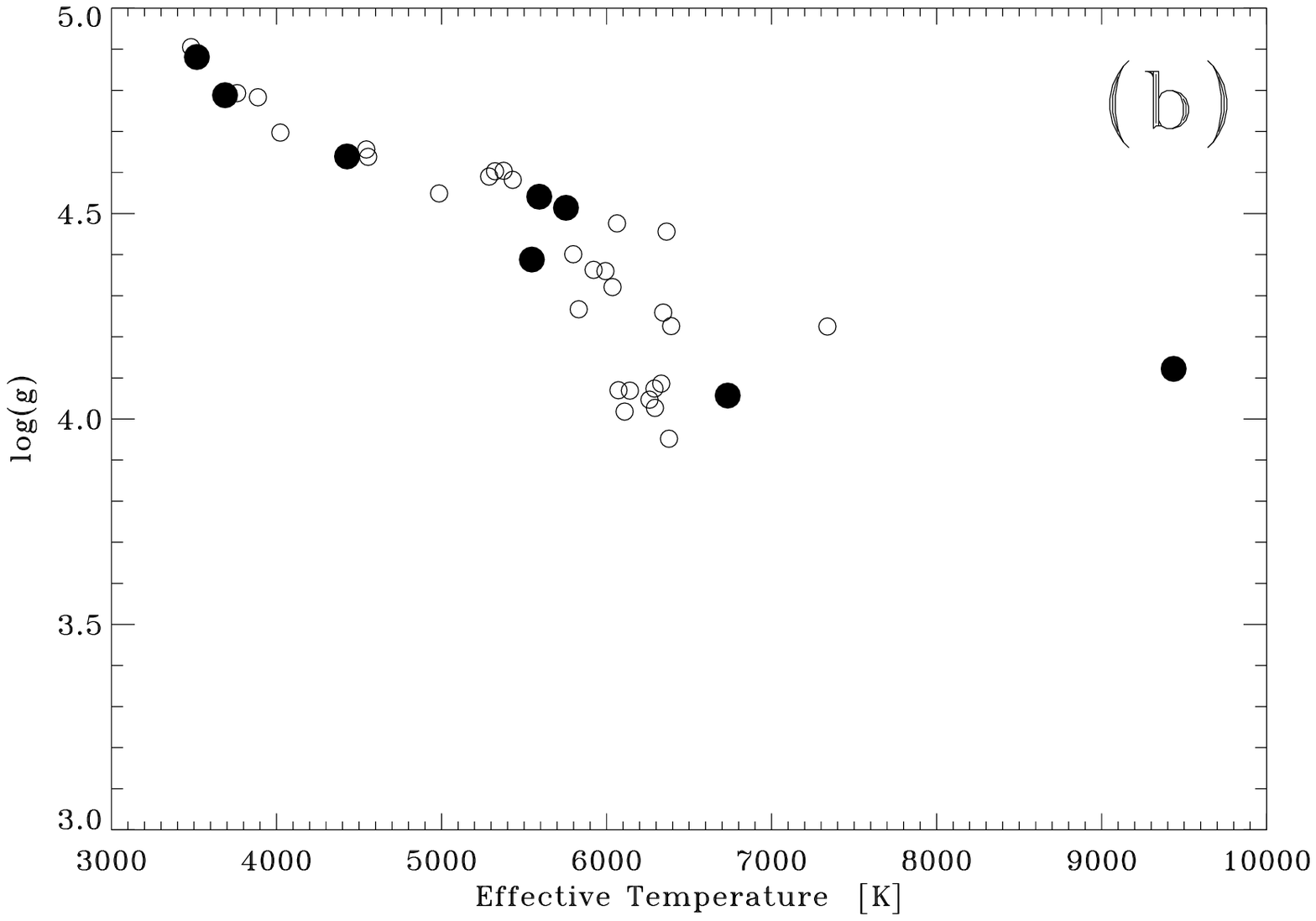}
\caption{
Surface gravity [log($g$)] as a function of effective temperature for 
the sample of
stars observed at (a) WIYN and (b) Gemini. Open circles represent stars
with no detection of a companion, and filled circles represent stars with 
detected companions.
}
\end{figure}

\begin{figure}[tb]
\plotone{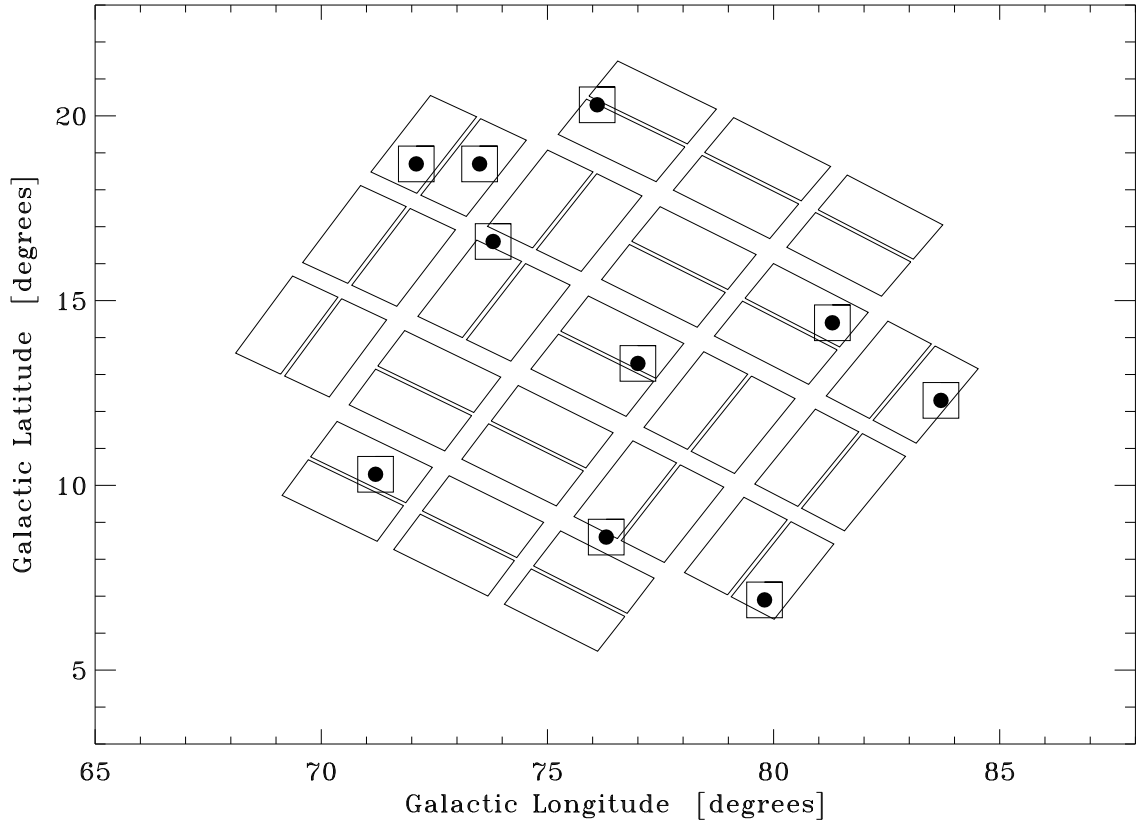}
\caption{
The ten pointings used for the TRILEGAL galaxy model in order to construct
the simulated samples discussed in the text. Each run of the TRILEGAL program
was a 1.0-square degree simulation, as indicated by the size of the box
around each point.
The outline of the {\it Kepler} CCDs is shown by rectangular shapes drawn
with solid lines.
}
\end{figure}

\begin{figure}[tb]
\plottwo{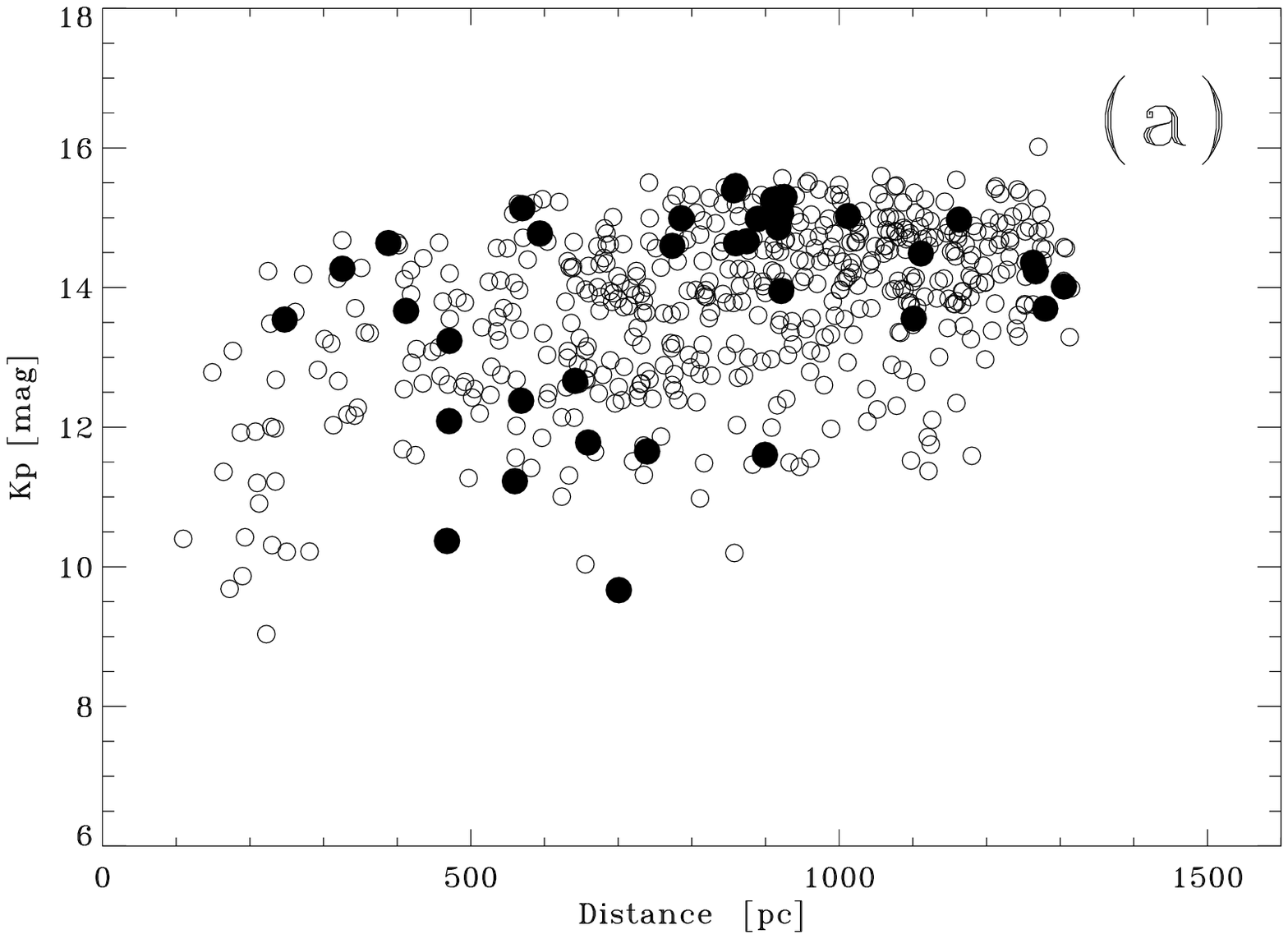}{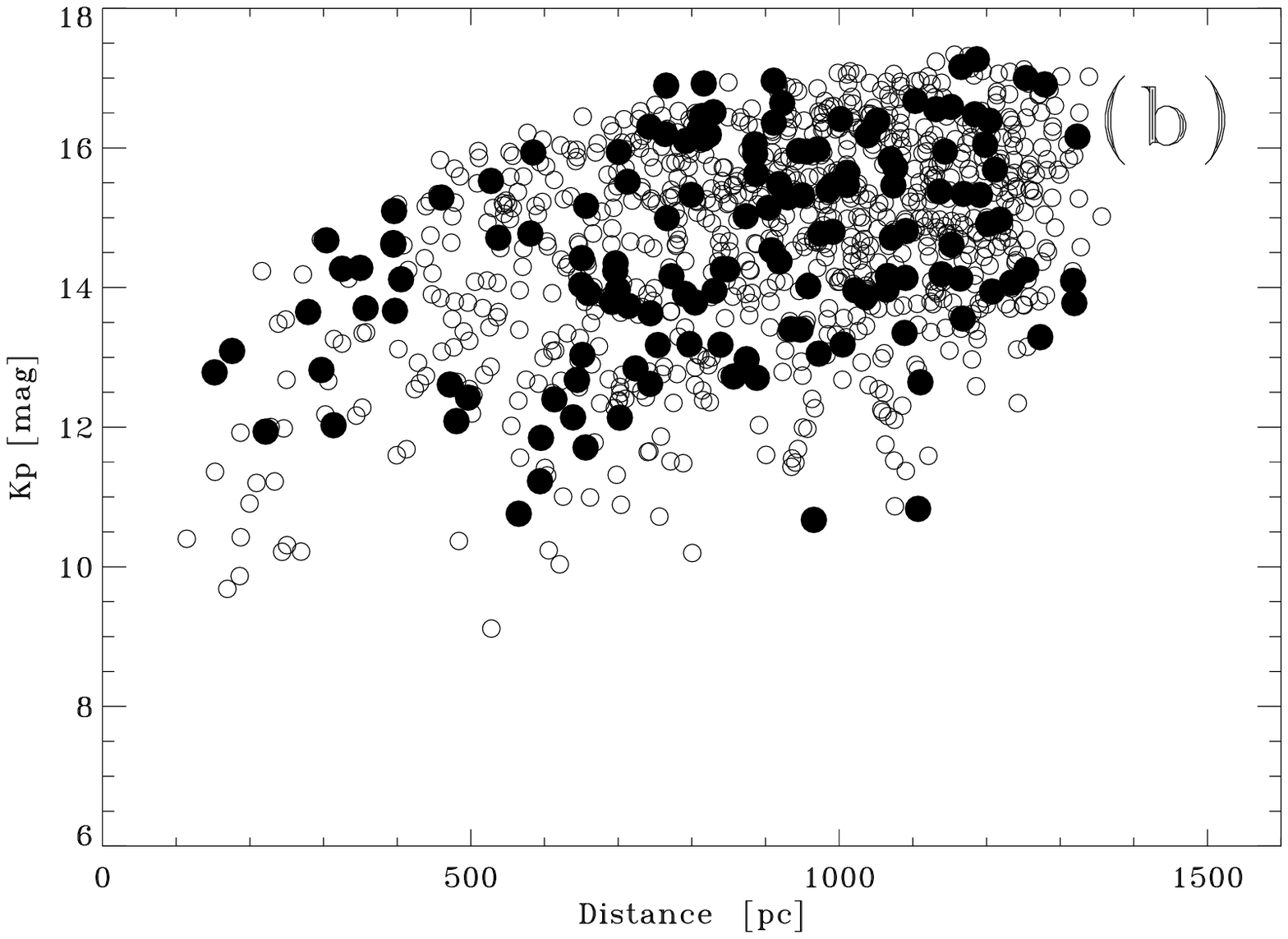}
\caption{
{\it Kepler} magnitude as a function of estimated distance for the simulation
results for (a) WIYN and (b) Gemini. Open circles represent stars
with no detection of a companion, and filled circles represent stars with 
detected companions.
}
\end{figure}

\begin{figure}[tb]
\plottwo{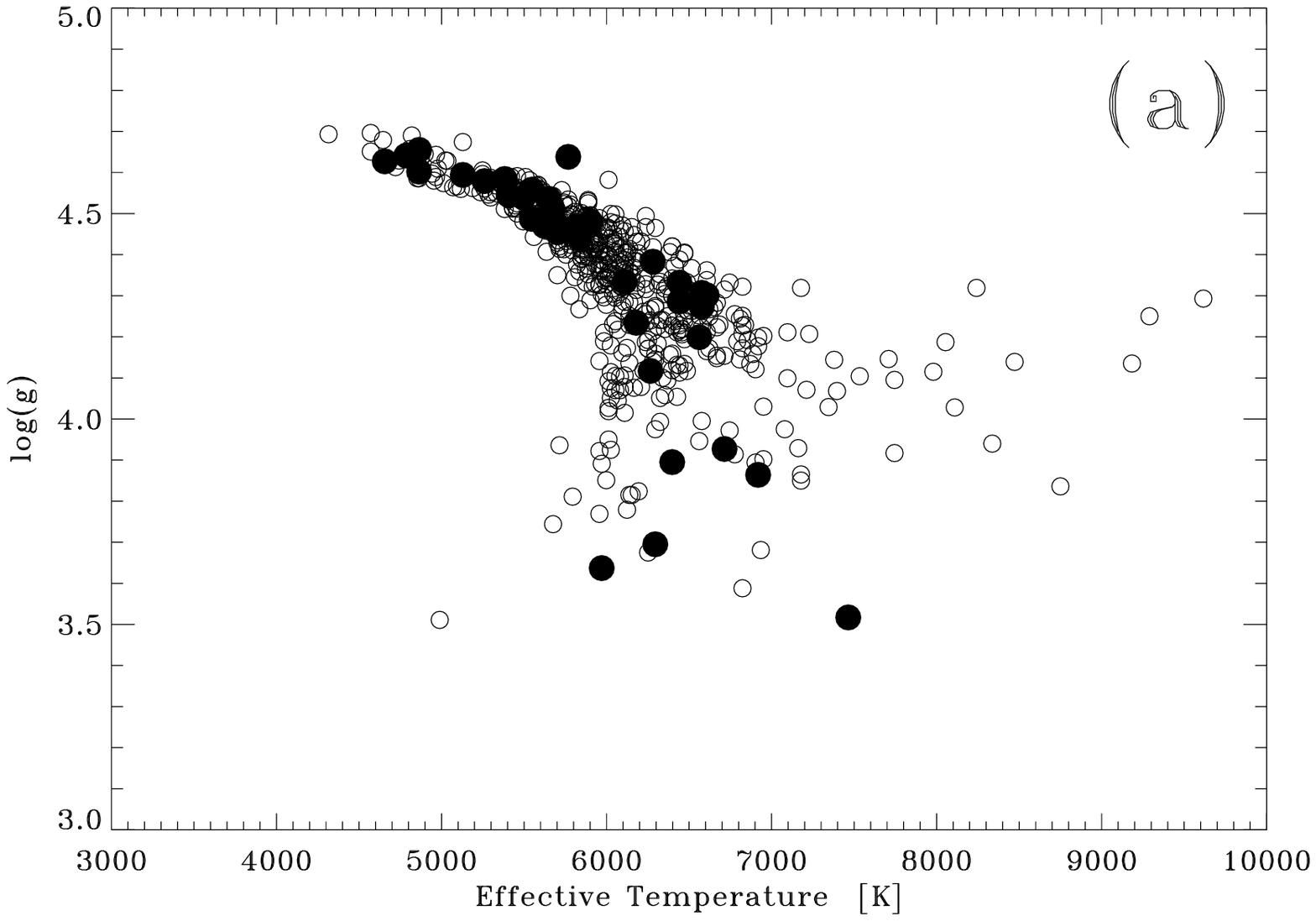}{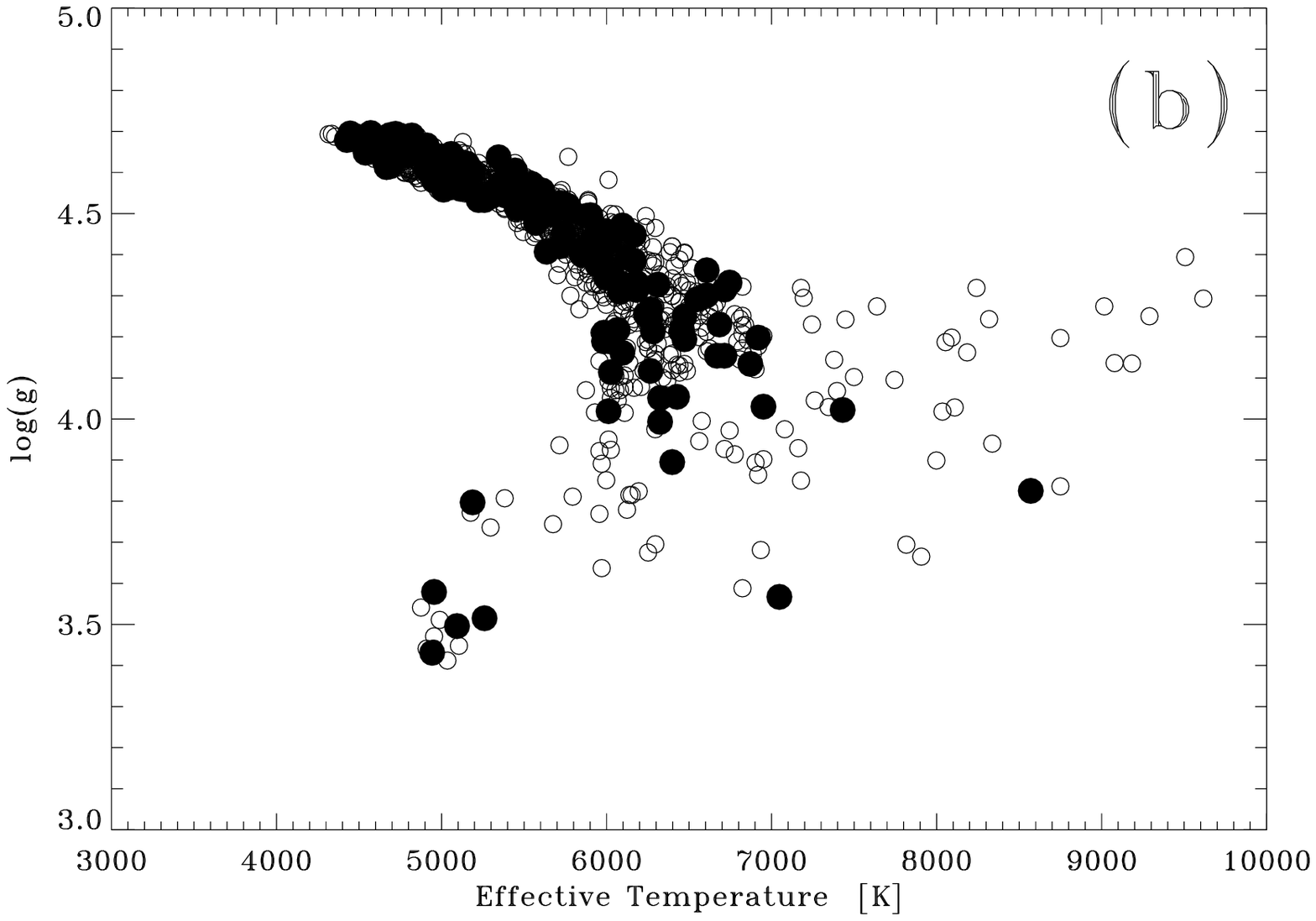}
\caption{
Surface gravity [log($g$)] as a function of effective temperature
for the simulation
results for (a) WIYN and (b) Gemini. Open circles represent stars
with no detection of a companion, and filled circles represent stars with 
detected companions.
}
\end{figure}

\begin{figure}[tb]
\plottwo{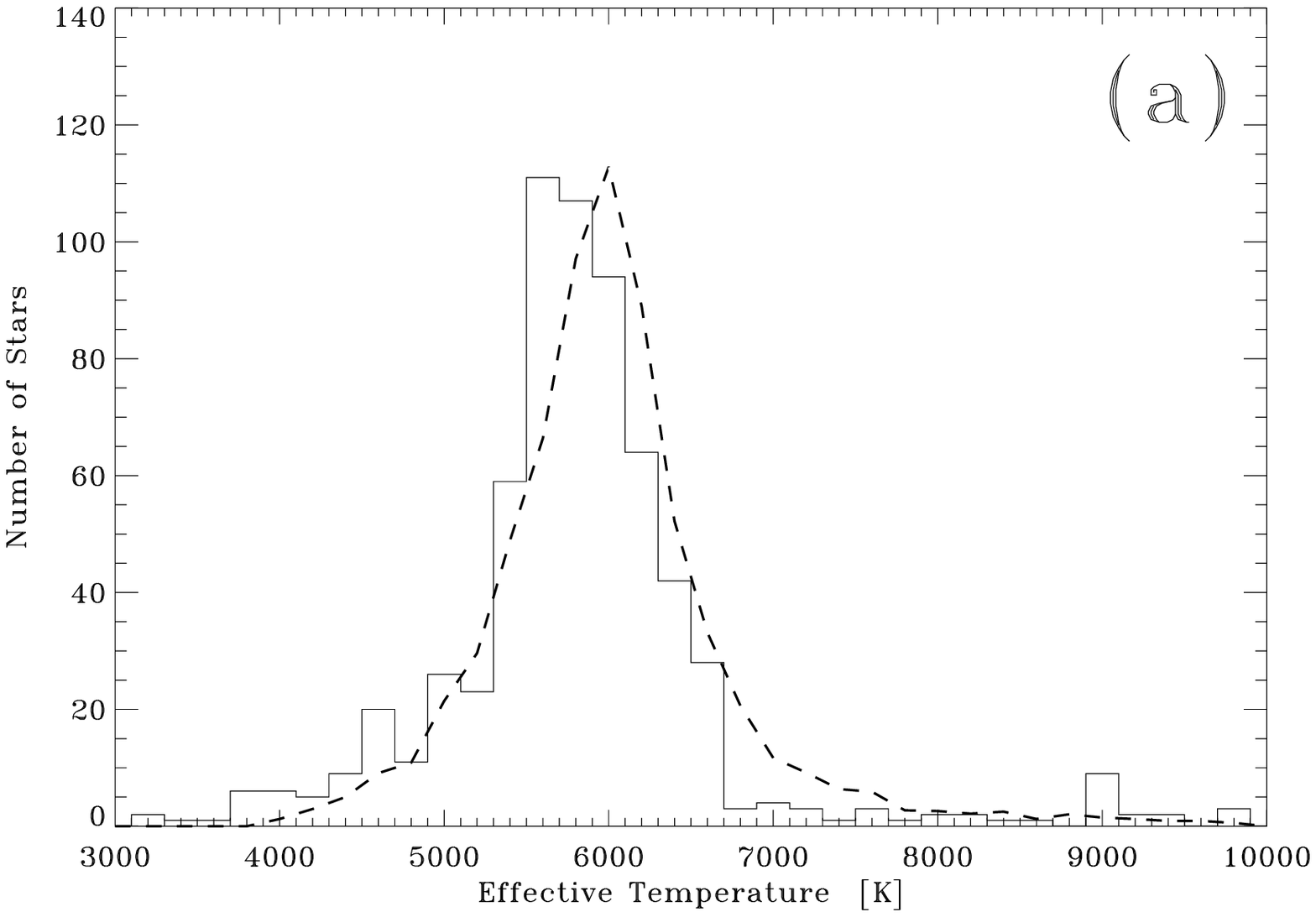}{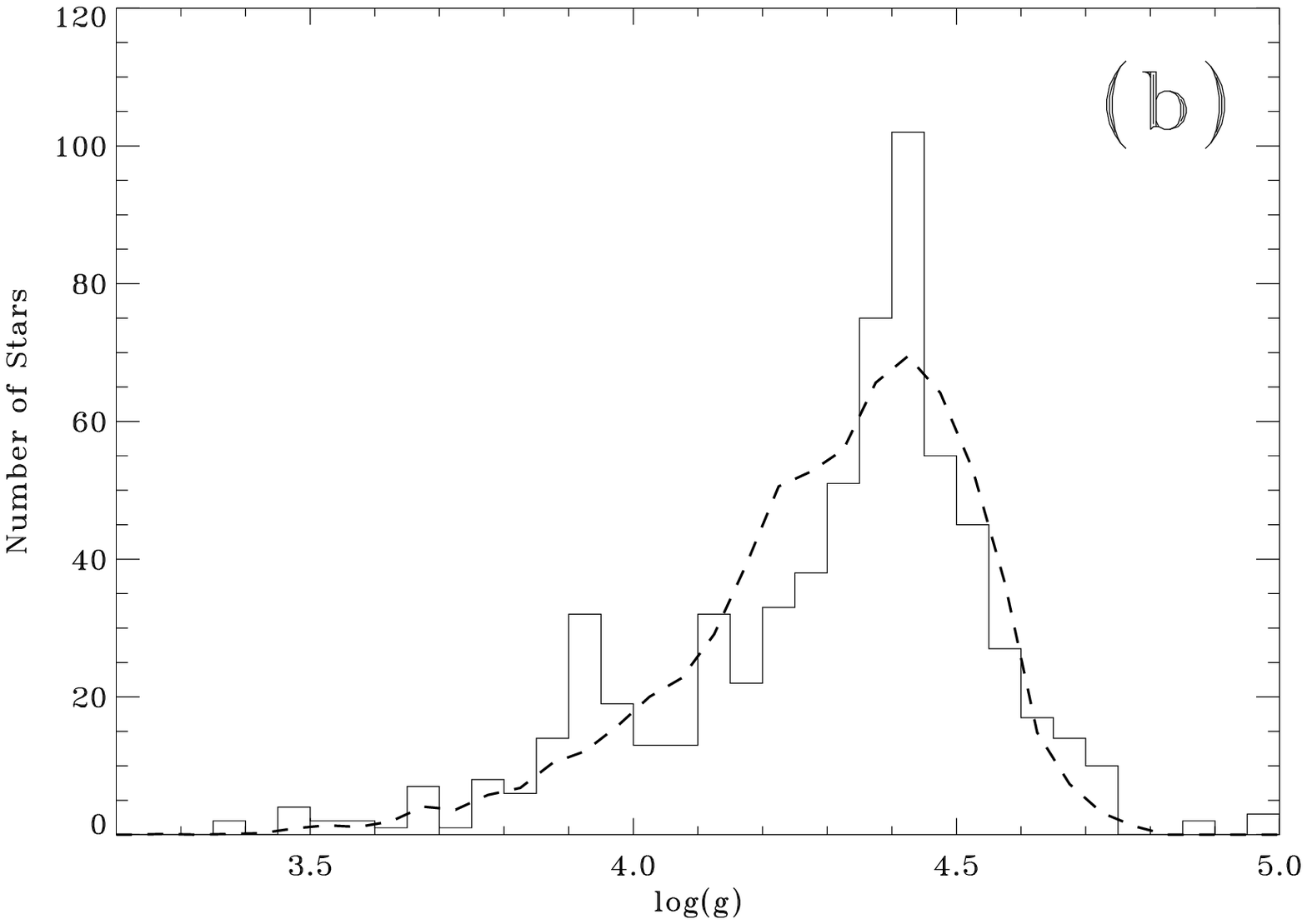}
\caption{
Histograms of (a) effective temperature and (b)
Surface gravity [log($g$)] for the observed WIYN sample of {\it Kepler} 
stars (solid
line)
versus the entire observable sample of simulated stars (dashed line).
The simulated results have been normalized to the total number of stars 
in the observed sample.
}
\end{figure}

\begin{figure}[tb]
\plotone{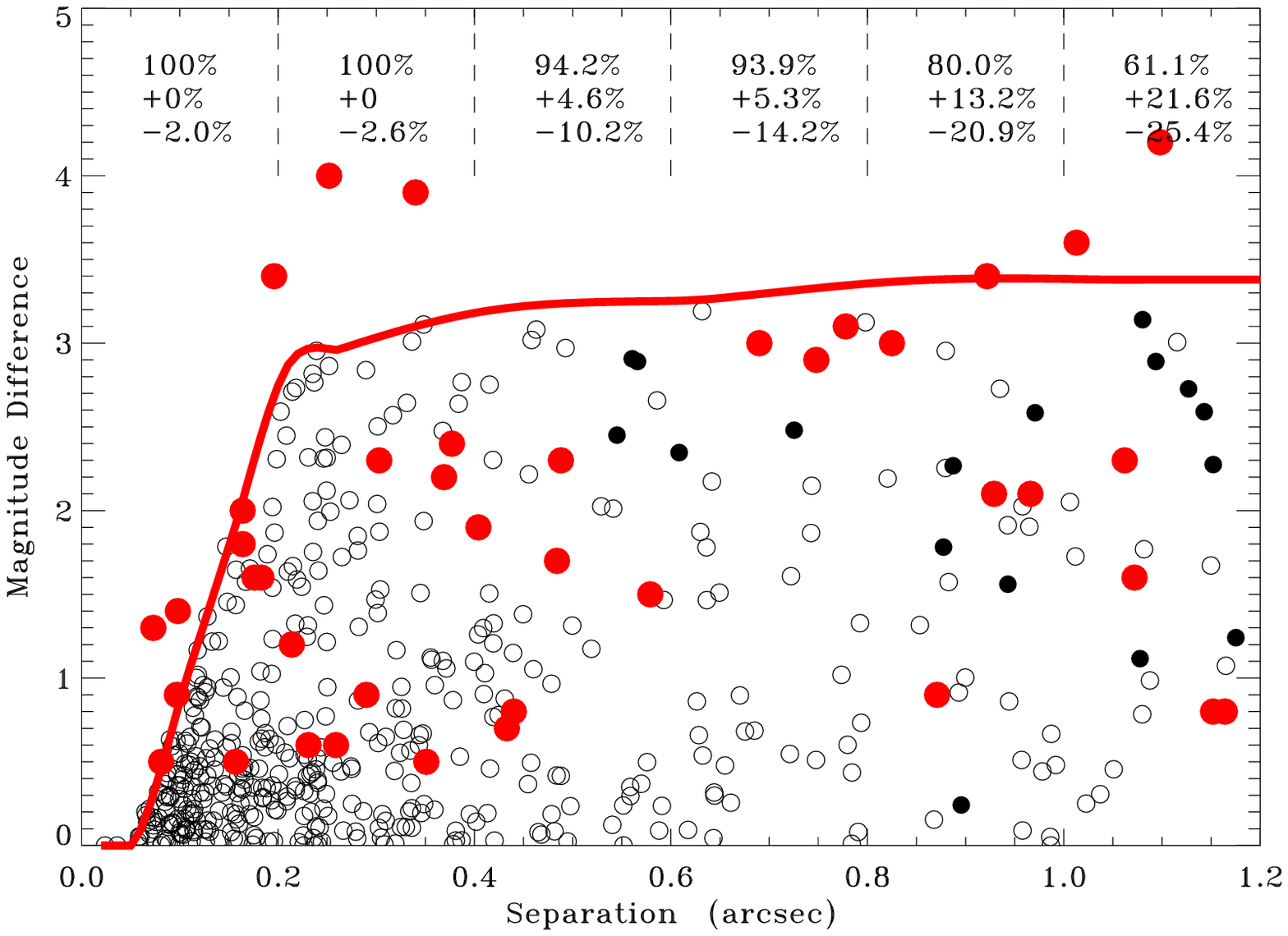}
\caption{
Simulation results for predicted stellar detections and observational results 
for WIYN at 692 nm. The open circles represent
detected bound components and the filled black circles represent 
detected line-of-sight components
from the simulation.
The filled red circles are
the locations on the diagram of components discovered at WIYN 
and the red curve is the 
average-quality detection curve appropriate for {\it Kepler} stars.
The numbers shown along the top of the diagram are the percentage of
bound companions for each 0.2--arcsecond-wide bin in separation.
}
\end{figure}

\begin{figure}[tb]
\plotone{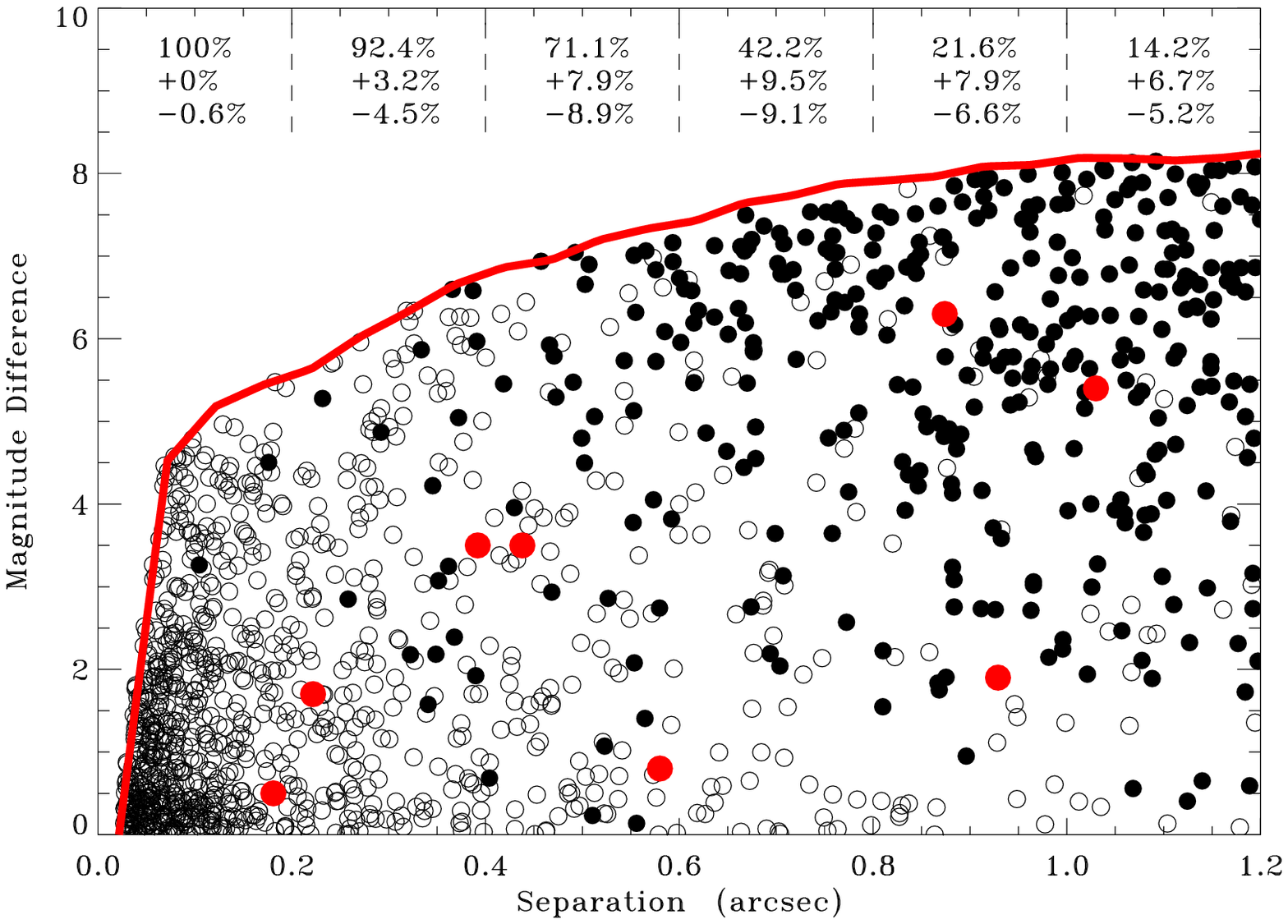}
\caption{
Simulation for predicted stellar detections and observational results 
for Gemini at 692 nm. The open circles represent detected
bound components and the filled black circles represent
detected line-of-sight components
from the simulation.
The filled red 
circles are
the locations on the diagram of components discovered at Gemini 
and the red curve is the 
average-quality detection curve appropriate for {\it Kepler} stars.
The numbers shown along the top of the diagram are the percentage of
bound companions for each 0.2--arcsecond-wide bin in separation.
}
\end{figure}

\begin{figure}[tb]
\plotone{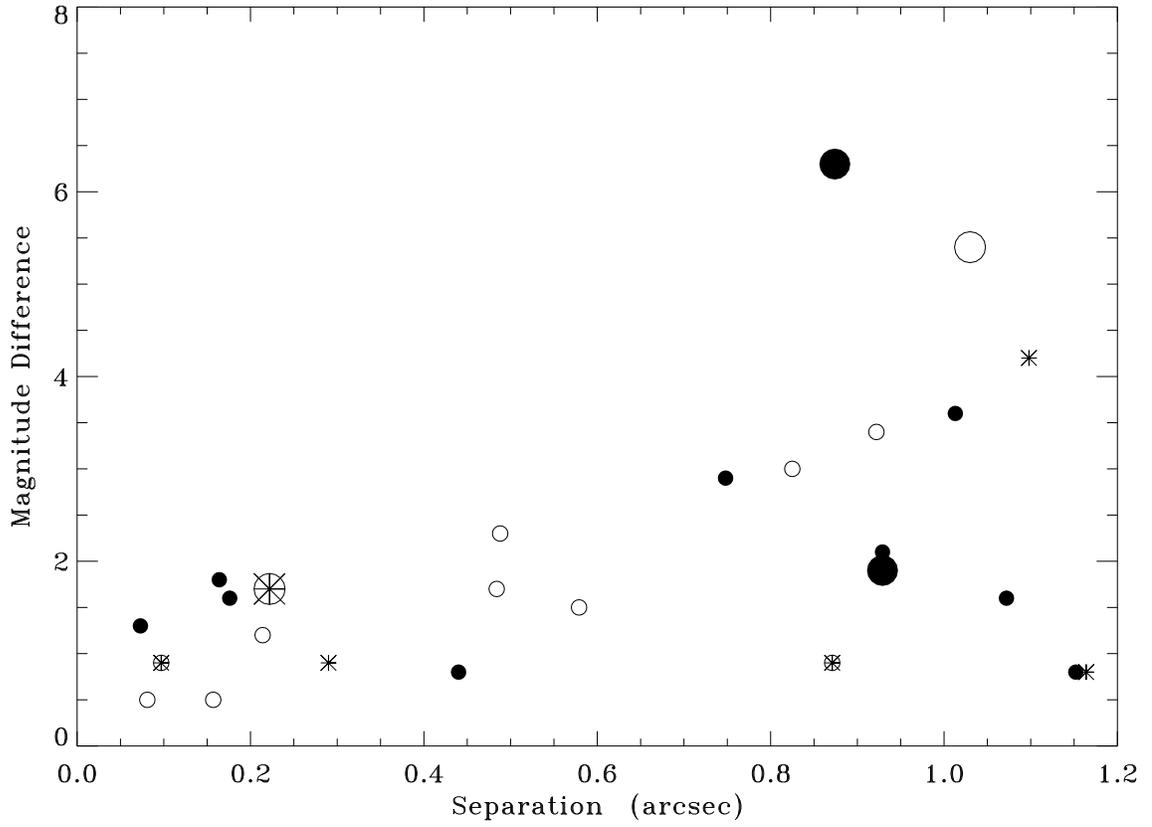}
\caption{
Magnitude difference versus separation for KOI speckle double stars 
that have been subsequently identified as false positives (filled
circles) and multi-planet candidate systems (open circles). 
Validated systems are shown with an asterisk symbol.
Objects observed at Gemini North are shown with plot symbols that are twice
as large as for WIYN data points.
}
\end{figure}

\clearpage

\begin{figure}[tb]
\includegraphics[scale=0.66,angle=90]{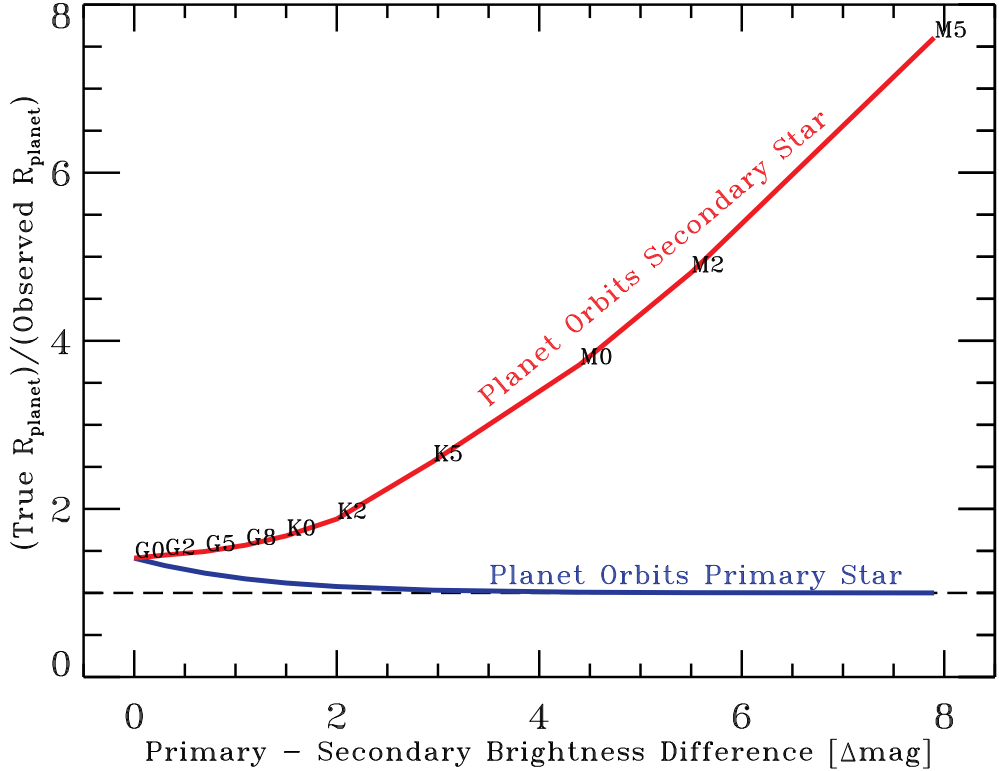}
\caption{
The ratio of the true planet radius to the observed planet radius for a
blended binary as a function of {\it Kepler} magnitude difference between the
companions.  The blue line represents the
ratio if the planet orbits the primary star; this line is the same for
all stars regardless of spectral type because the derived radius is only
affected by the transit dilution of the secondary star ({\it i.e.\ }the
presumed stellar radius does not change significantly).  The red line
represents the ratio if the planet orbits the secondary star and takes
into account the brightness deblending and the difference in the stellar
radii between the primary and secondary stars. The positions of the
possible secondary stars are marked in black. The exact shape of this
line is dependent upon the primary stellar type as the secondary star,
by definition,
will be fainter and smaller than the primary. In this example,
the primary star is a G0V. 
}
\end{figure}

\end{document}